\documentclass[showpacs,prb,superscriptaddress,aps,floatfix]{revtex4-2}

\usepackage[english]{babel}
\makeatletter\AtBeginDocument{\let\@elt\relax}\makeatother
\usepackage[utf8x]{inputenc}
\usepackage{graphicx}
\usepackage{amsmath}
\usepackage{amssymb}
\usepackage{braket}
\usepackage{hyperref}
\usepackage{makeidx}
\usepackage{color}
\usepackage{ulem}
\usepackage{nicefrac}
\usepackage{bm}

\begin{document}

\title{Interlayer and intralayer excitons in AlN/WS$_2$ heterostructure}

\newcommand{\cinam}{CNRS/Aix-Marseille Universit\'e, Centre Interdisciplinaire de Nanoscience de Marseille UMR 7325 Campus de Luminy, 13288 Marseille cedex 9, France}
\newcommand{\torvergata}{Dipartimento di Fisica, Universit\'a  di Roma Tor Vergata, and INFN, Via della Ricerca Scientifica 1, I-00133 Rome, Italy}
\newcommand{\lem}{Laboratoire d'Etude des Microstructures, ONERA-CNRS, BP 72, 92322 Ch{\^a}tillon Cedex, France}
\newcommand{\piim}{Universit\'e Aix-Marseille, Laboratoire de Physique des Interactions Ioniques et Moléculaires (PIIM), UMR CNRS 7345, F-13397 Marseille, France}
\newcommand{\etsf}{European Theoretical Spectroscopy Facilities (ETSF)}
\newcommand{\cambridge}{Cavendish Laboratory, University of Cambridge, J.\,J.\,Thomson Avenue, Cambridge CB3 0HE, United Kingdom}

\graphicspath{{./}}

\author{C. Attaccalite}
\affiliation{\cinam}
\affiliation{\etsf}
\author{M.S. Prete} 
\affiliation{\etsf}
\author{M. Palummo}
\affiliation{\torvergata}
\affiliation{\etsf}
\author{O. Pulci}
\affiliation{\torvergata}
\affiliation{\etsf}

\date{\today}

\begin{abstract}
The study of intra and interlayer excitons in 2D semiconducting vdW heterostructures is a very hot topic not only from fundamental but also applicative point of view. Due to their strong light-matter interaction, Transition Metal Dichalcogenides (TMD) and group-III nitrides are particularly attractive in the field of opto-electronic applications such as photo-catalytic and photo-voltaic ultra thin and flexible devices. Using first-principles ground and excited-state simulations, we investigate here the electronic and excitonic properties of a representative  nitride/TMD heterobilayer, the AlN/WS$_2$. We demonstrate that the band alignment is of type I and low energy intralayer excitons are similar to those of a pristine WS$_2$ monolayer. Further, we disentangle the role of strain and AlN dielectric screening on the electronic and optical gaps.
These results, while on one side do not favor the possible use of AlNWS$_2$ in photo-catalysis, as instead envisaged in the previous literature, can boost the recently started experimental studies of 2D hexagonal Aluminum nitride as a good low screening substrate for TMD-based electronic and opto-electronic devices.
Importantly, our work shows how the inclusion of both spin-orbit and many-body interactions is compulsory for a correct prediction of the electronic and optical properties of  TMD/nitride heterobilayers.
\end{abstract}

\maketitle


\section{Introduction}

Guided by graphene rise\cite{geim2010rise}, a broad family of two-dimensional (2D) materials 
with different electronic and optical properties\cite{ponraj2016photonics} is currently studied for fundamental research and also for a plethora of envisaged device-oriented applications.\cite{lin20162d} 
Due to their flat nature, 2D materials show unique potential to fabricate flexible and ultra-thin devices with the hope to reduce costs production and improve performances. Layered heterostructures offer a unique playground to engineer their electronic and optical properties\cite{Li_2016,Brozzesi2022}, thanks to the availability of metallic, semiconducting and insulating monolayers (MLs) and to the possibility to stack them  with any order and orientation using Van der Waals growth\cite{growth2D}. Hence, there is  essentially no epitaxial lattice-match requirements typical of 3D materials.

Two emerging classes of 2D materials, Transition Metal Dichalcogenides (TMDs) and group-III nitrides,  are particularly attractive for their sizable band-gaps, strong light-matter interaction and their interesting excitonic properties which hold premise for their use in opto-electronic applications,
such as ultra-thin and flexible photovoltaic (PV) or photo-catalytic cells and light-emitting diodes (LED). 

After the discovery of indirect to direct gap behavior reducing the thickness of MoS$_2$ to  monolayer form  \cite{Splendidiani_2010,Mak_2010}, the structural, electronic and optical properties of two-dimensional TMDs have been widely investigated at theoretical level by means of   Density Functional Theory (DFT)  and refined excited state methods, (namely GW and BSE). \cite{Rasmussen2015,chen2018theory,Bernardi_2017,Gusakova2017} 
At the same time, an increasing number of DFT studies\cite{kecik2018fundamentals,Wang2021} has been published for 2D nitrides  and several predictions based on GW and BSE methods exist in the literature\cite{Sahin.Cahangirov.ea:2009:PRB,prete2017tunable, Prete.Pulci.ea:2018:PRB,prete2020giant}.
TMD vdW heterostructures (HTs) have been widely investigated in recent years, and also some studies appeared in the literature on their combination with other 2D materials.\cite{mohanta2019interfacing,wang2021electronic,rawat2021interfacing} However much less attention has been dedicated to HTs obtained combining 2D TMDs and nitrides.

In particular, it is of interest to identify if a type I or type II band alignment is present and how the electronic and optical properties of an isolated TMD change in presence of a nitride (beyond the normally used hBN) substrate. 

Indeed, while a type I band alignment can be a  good prerequisite for strongly bound excitons associated to efficient light-emission, a type II favors the formation of long-lived interlayer excitons and enable their ultra-fast charge transfer \cite{Jiang2021}.
This is interesting not only for the investigation of  novel excitonic physics, such as quantum Bose gases\cite{Varsano2020,Sun2022,Wang2021}  but also for the design of innovative opto-electronic devices.

Then, motivated also by a recent experimental work \cite{Chang2022} which show how 2D hexagonal AlN can grown by using ALD on TMD, resulting a gate dielectric material alternative to h-BN,\cite{prete2020giant} and by other works \cite{Song2021,Desai2020,Yan2018}
showing the growth of hexagonal group III-nitrides on TMDs, our goal here is to investigate the structural and opto-electronic properties of a prototype bi-layer composed of AlN/WS$_2$. 
Our aim is to focus  on the role of many-body effects, which are expected to be of primary importance due to the reduced dimensionality and to the low dielectric screening.
It is worth to mention that in a previous study Liao and collaborators \cite{liao2014design} proposed  this class of heterobilayers 
for next generation of ultra-thin flexible opto-electronic devices
(with a focus on AlN(GaN)/MoS$_2$), but remain at single-particle DFT level of approximation.

\section{Methods and computational Details}
Our first-principles calculations are based on DFT and many body perturbation theory (MBPT).
For DFT structural calculations we have used PBEsol   exchange-correlation functional\cite{perdew2008restoring} and DOJO pseudopotentials\cite{van2018pseudodojo} within the {\it Quantum Espresso} code \cite{QE}.  Van der Waals corrections \cite{hamada2014van} are applied on top of the PBEsol functional in order to take into account the weak interaction between the layers in the AlN/WS$_2$ heterostructure.
A ($24\times24\times1$) ${\bf k}$-point mesh and an energy cutoff of 120 Ry have been used. 
In order to avoid spurious interaction between adjacent images, a supercell of 35.5 Bohr thickness is used and a cutoff on the Coulomb interaction is applied.\cite{rozzi2006}
The electronic band structure and optical absorbance spectra are  calculated taking into account spin-orbit coupling (SOC), both at the DFT level of approximation and beyond.
The MBPT simulations  are performed  using the Yambo code \cite{sangalli2019many}, namely we first  corrected the DFT band structure by doing $G_0W_0$ calculations\cite{aryasetiawan1998gw} and  then solved the Bethe Salpeter equation (BSE)\cite{strinati1988application} to obtain the optical properties taking into account excitonic and local-field effects as well as the full spinorial nature of the electronic wavefunctions \cite{Marsili_prb2021}.
The parameters of convergence are the following: 90~Ry cutoff for the wavefunctions, a ($33 \times 33 \times 1$) {\bf k}-point mesh, both for the exchange and correlation part of the self-energy, 5 Hartree cutoff for the dielectric matrix and 500 bands. In order to speed up convergence with the number of empty bands, we use the a terminator both in G and W.\cite{bruneval2008accurate} 
Note that for W  we used the Godby-Needs plasmon pole model\cite{godby1989metal}, which does not based on sum-rules, and therefore it requires a lower number of conduction bands to converge.\cite{stankovski2011g}
For the calculation of the BSE optical spectrum we have included the 8 highest valence bands and the 8 lowest conduction bands, and the same k-point sampling of the G$_0$W$_0$ calculations.
\section{Results}

\subsection{Crystal and Electronic Structures of AlN/WS$_2$ vdW Heterostructure}

\begin{figure*}[!ht]
       \centering
       \includegraphics[width=1.0\textwidth]{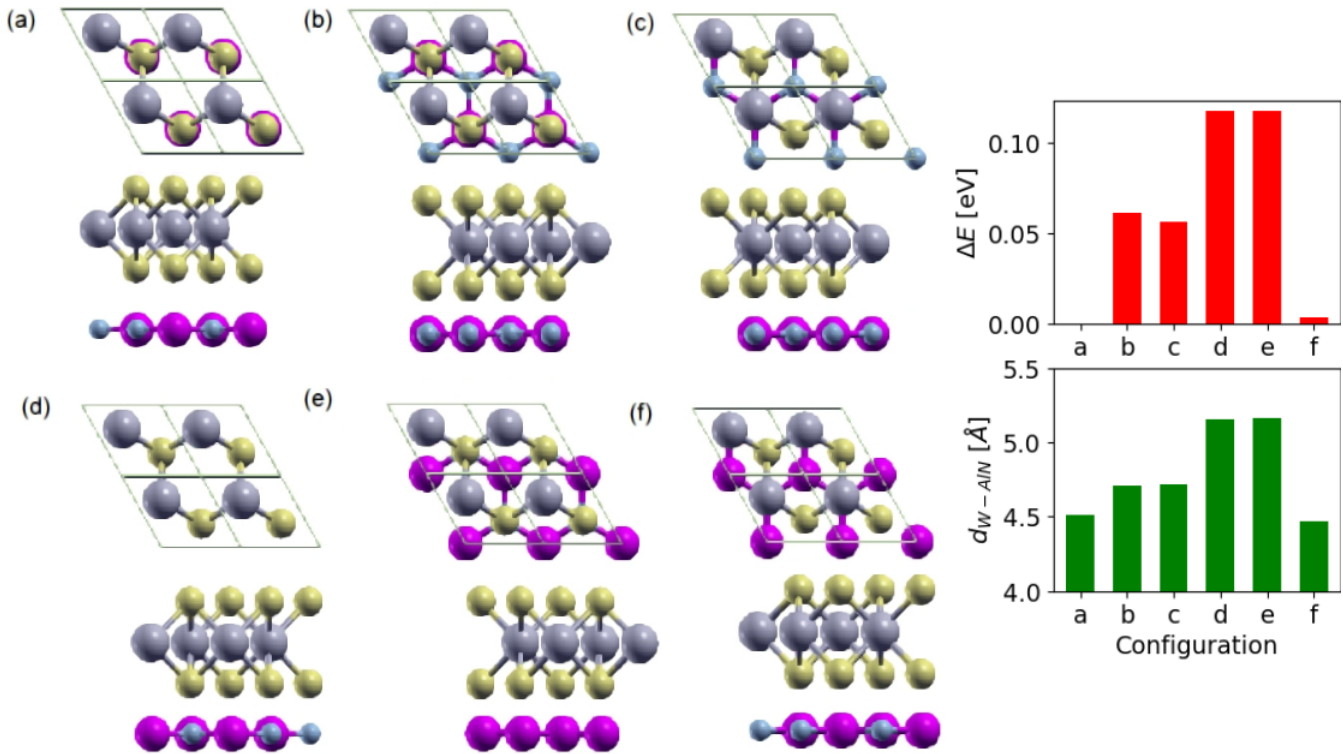}
	     \caption{Top and side views of AlN/WS$_2$ heterostructures in different rotation angles: (a) 0$^\circ$, (b) 60$^\circ$,(c) 120$^\circ$, (d) 180$^\circ$, (e) 240$^\circ$ and (f) 300$^\circ$. Small light-blue balls  indicate N atoms; in violet Al atoms; in gray and yellow are indicated W and S atoms, respectively. On the right we report the energy of each configuration respect to the $(a)$ one and the distance $d_{W-AlN}$.} \label{fig:configurations}
      \end{figure*}

      The relaxed lattice parameters are 
      3.11 \r{A} and 3.16 \r{A} for isolated AlN and WS$_2$ monolayers, respectively. The lattice mismatch  is less than 2\%, which is a good value for constructing AlN/WS$_2$ heterostructure based on a 1$\times$1 cell-periodicity. As starting point we chose a lattice parameter intermediate between those of the layers,   then we relax the combined structure  obtaining a final lattice parameter  of 3.142 \r{A}. 
      For this HT we consider six 
      configurations, as illustrated in Fig.\ref{fig:configurations}, where 
      different rotation angles and stacking between the adjacent sheets have been selected. The cell and the atomic positions for all configurations are then relaxed using a $24 \times 24 \times 1$ k-point sampling, 120~Ry cutoff for the wave-function and a cutoff on forces of $10^{-4}$~Ry/au.
  
      \begin{table}[!h]
      \centering
      \begin{tabular}{|c|c|c|c|c|c|}
      \hline \hline
	      conf. & $\Delta$E(eV) & $a$(\AA) &  $d_{W-AlN}$(\AA) & L$_{W-S}$(\AA) & L$_{Al-N}$(\AA) \\ \hline
	      a        & 0             &  3.142 &  4.509           & 2.405     &  1.814     \\ \hline
	      b        & 0.0613  &  3.140  &     4.705       & 2.405     & 1.813       \\ \hline
	      c        & 0.0566  &  3.138  &     4.717     & 2.406     &   1.812     \\ \hline
	      d        & 0.1176  &  3.137  &     5.158       &  2.404     &    1.811     \\ \hline           
	      e        & 0.1179  &  3.137  &     5.164       &  2.405     &  1.811     \\ \hline
	      f        & 0.0032  &  3.144  &     4.469      &   2.403  &    1.815    \\ \hline \hline
      \end{tabular}
	      \caption{Energy difference $\Delta$E (eV),  layer distance $d_{W-AlN}$,  L$_{W-S}$ and L$_{Al-N}$ bond lengths of the different configurations of AlN/WS$_2$  heterostructures calculated by DFT-PBEsol including vdW correction. Notice that for the L$_{W-S}$ distance we consider the S atom towards the AlN layer.}
      \label{tab:AlN/WS_2_groundstate}
      \end{table}
    
      In Table \ref{tab:AlN/WS_2_groundstate} we list the calculated energy difference between the total energy of various stacking configurations and the most stable one, the interlayer distance  between AlN and WS$_2$ layers, as well as the W-S and Al-N bond lengths for the AlN/WS$_2$ heterostructures.
      The energy difference $\Delta E$ is defined as $\Delta E=E-E_0$ where E$_0$ is the total energy of the most stable configuration and E is the total energy of each configuration. The calculated most stable structure (with $\Delta$E = 0 eV) for AlN/WS$_2$ is the {\bf (a)} configuration, where N(Al) atoms lay on top of W(S) (see Fig.\ref{fig:configurations}). 
      In order to investigate the thermodynamic stability of this configuration, we calculate a stacking energy  value of -12.45 meV, according to 

      \begin{equation}
      E_{stack}=E_{total}^{AlN/WS_2}-E_{total}^{WS_2}-E_{total}^{AlN}
      \end{equation}
     \vspace*{0.1cm}
            where E$_{total}^{AlN/WS_2}$, E$_{total}^{WS_2}$ and E$_{total}^{AlN}$ represent the total energies of AlN/WS$_{2}$ heterostructure, of WS$_2$, and of AlN monolayers, respectively. The negative value indicates that this heterostructure is energetically favorable and could be experimentally realized. 

       \begin{figure}[!ht]
        \centering
        \includegraphics[width=0.49\textwidth]{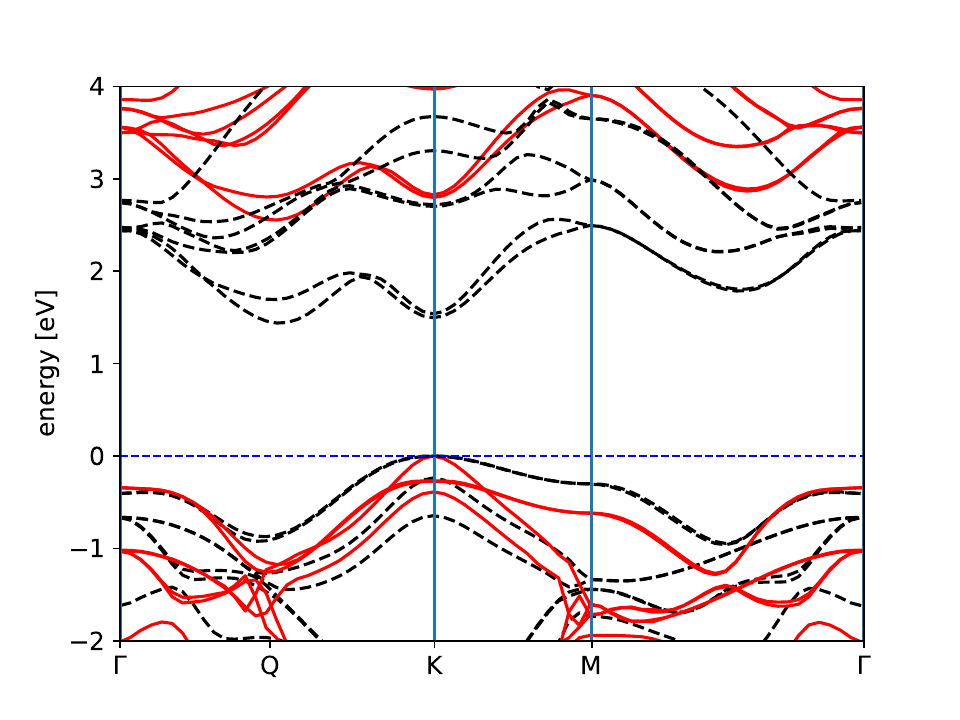}
	      \caption{Band structure for the relaxed AlN/WS$_2$ vdW heterostructure along the high symmetry path of the first Brillouin zone calculated within DFT (dashed black) and G$_0$W$_0$ (continuous red) level of approximation, taking into account spin-orbit interaction. Notice that the minimum of the conduction bands appears between $\Gamma$ and $\bf K$, close to the $\bf Q$ point, and together with the top of the valence band at $\bf K$ forms the indirect band gap. The zero energy is set at the top of the valence bands in both the DFT and G$_0$W$_0$ calculations.} \label{fig:AlN/WS2_Bands}
      \end{figure}
  \begin{table}[!h]
      \centering
      \begin{tabular}{|c|c|c|c|}
      \hline \hline
	      System   & Dir. Gap(eV) & Ind. Gap(eV) & E$_b$ \\ \hline
	       AlN  (a=3.11 \AA)   &  6.27(3.65)   &  5.61(2.97)  &  1.79 \\ \hline
	       WS$_2$ (a=3.16 \AA) &  2.76(1.64)   &  2.60(1.64)  &  0.64 \\ \hline 
	       AlN/WS$_2$ (a=3.142 \AA)&  2.80(1.50) &   2.56(1.44) & 0.59  \\ \hline \hline
	         AlN/WS$_2$ (a=3.16 \AA) &  2.71(1.38) &   2.56(1.38) & 0.59   \\ \hline \hline
      \end{tabular}
	      \caption{Direct {\bf K-K} and indirect {\bf K-Q} band gaps for the three system at the G$_0$W$_0$ level. DFT results are given in parenthesis. $E_b$ is the binding energy of the lowest bright exciton. The 3rd row of the table refers to the relaxed bilayer, whereas the last row refers to the AlN/WS$_2$ HT forced to keep the same lattice parameter of the isolated WS$_2$.}
      \label{GW_table}
      \end{table}

      For isolated WS$_{2}$ and AlN our electronic DFT results are consistent with existing literature\cite{Strain_TMD_Deng,Shi_WS2_2013,PrB_TMD_Rohlfing}.
       It is well known that the direct/indirect gap nature of TMDs monolayers  is strongly dependent on the lattice parameters and small changes in the approximations used in the DFT calculations. For this reason in this work we used a modified GGA exchange
functional\cite{hamada2014van}, together with nonlocal correlation for the second version of the van der Waals density functional of Lee et al.\cite{lee2010higher}. This functional has shown very good performance in two-dimensional heterostructures\cite{hamada2014van}.\\ 
       WS$_{2}$ has an direct DFT gap of 1.64 eV located at the {\bf K} point with a spin-orbit splitting of 0.036~eV for the first conduction bands.  
    The AlN monolayer shows an indirect DFT gap of 2.97~eV with CBM and VBM located at $\bm \Gamma$ and $\bf K$ points, respectively,  in agreement with previous calculations.\cite{prete2017tunable} 
    \par
    When we combine the two systems, forming the AlN/WS$_{2}$ heterostructure, the final DFT electronic bands present an indirect gap  of E$_{gap}^{KQ}$=1.44 eV, between ${\bf K}$ and {\bf Q}, while the direct one is located at $\bf K$, E$_{gap}^{KK}$=1.50 eV (see 3rd row of table \ref{GW_table} and Fig.\ref{fig:AlN/WS2_Bands}).
These gaps are smaller than those of the separated layers and although could be quite unexpected for a vdW heterostructure, it has been reported several times in the literature \cite{bernardi2013extraordinary,Rawat2019,Refiore_2020,Bastonero_2021} and suggests a type II junction.
      
Indeed, calculating the projected density of states (PDOS), 
it  results that, near the  
the CBM and VBM states
are localized on different monolayers of the vdW heterostructure (see Fig.1 in the Supplementary Material). 

      

\begin{figure}[!ht]
        \centering
        \includegraphics[width=0.4\textwidth]{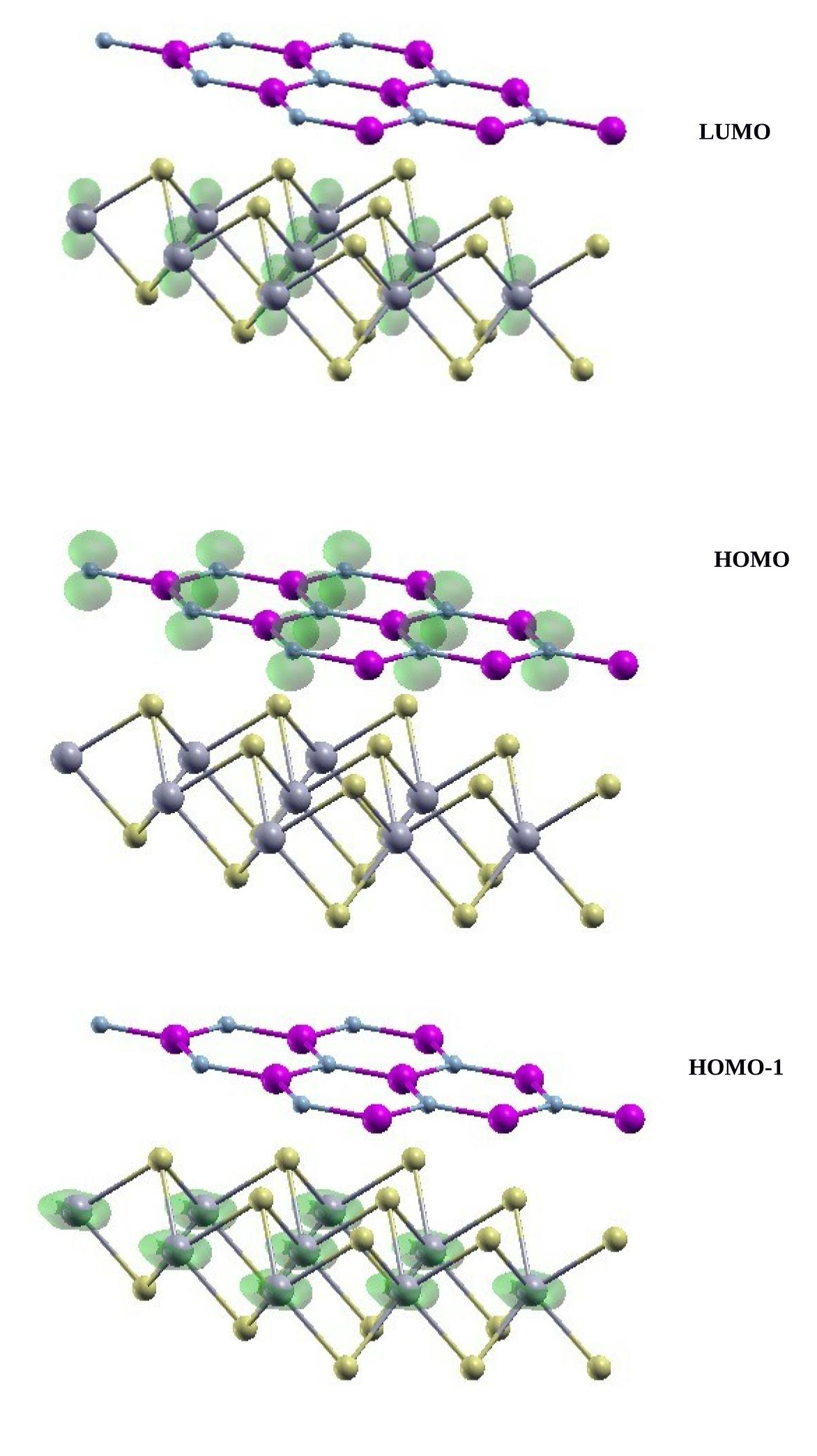}
	      \caption{Band-decomposed charge density of the LUMO (top) and HOMO (middle) and HOMO-1 (bottom) at the DFT level for AlN/WS$_{2}$ vdW heterostructure at $\bf K$ point. The LUMO at $\bf Q$, not reported the figure, is very similar to the LUMO at $\bf K$.} \label{fig:AlN/WS2_charge}
\end{figure}

 As seen more clearly
in Fig. \ref{fig:AlN/WS2_charge}, the HOMO, located at {\bf K},   mainly consists of the N $p_{z}$ orbitals while the LUMO, located at {\bf Q}, is mainly due to W $d_z$ orbitals.

The interaction between the two layers increases also the spin-orbit splitting of the conduction bands at {\bf K} to 0.042~eV with respect to the isolated WS$_2$. 
This is not unexpected because it has already been shown that charge transfer and strain can modify the spin-orbit splitting of two dimensional materials.\cite{PhysRevB.94.115131,doi:10.1063/1.5096413} 

\par\noindent
For what concerns DFT level of approximation, we can then affirm that  the AlN/WS$_2$ is a type II heterostructure, hence suggesting it could have promising properties for charge separation in photo-voltaic/photo-catalytic devices.\cite{liao2014design}\\ 
In order to confirm this interesting result, 
     we need to go beyond DFT to better take into account exchange and correlation effects. We hence applied the perturbative one-shot GW method to AlN/WS$_2$ and to the separated layers. It is worth to remember that in this work, differently from ref.~\cite{liao2014design}, SOC are included both in DFT and in MBPT calculations \cite{Marsili_prb2021}.\\
 \subsection{Quasi-particle effects}    
 We report in Table \ref{GW_table} the direct and indirect gap at the different levels of approximation of all the three systems AlN, WS$_2$, and AlN/WS$_2$. We discuss here the results for the relaxed AlN/WS$_2$ structure (3rd row in table \ref{GW_table}). While the constrained structure (last row in table \ref{GW_table}) will be discussed in section \ref{subsec:strain}.

 The GW correction in AlN opens the gap up to 5.61~eV without changing its position in $k$ space. 
 In WS$_2$, instead, the calculated GW corrections open the gap and  make WS$_{2}$ an indirect gap material with the minimum gap located between {\bf K}=$(\nicefrac{1}{3},\nicefrac{1}{3},0)$ and {\bf Q}=$(\nicefrac{10}{66},\nicefrac{10}{55},0)$   \cite{Shi2013}. 
  The GW also reduces the  spin-orbit splitting of the lowest conduction bands at {\bf K} to 0.016~eV.\\
Now we move the AlN/WS$_2$ heterostructure. The accurate GW electronic band structure of the relaxed bilayer is displayed in Fig. \ref{fig:AlN/WS2_Bands}. The inclusion of the GW corrections reduces the SOC splitting at {\bf K} to 0.028~eV. Interestingly, the quasi-particle corrections induce an interchange of the bands close to the $\bf K$ point. The DFT highest occupied double degenerate valence bands (HOMO), located on the nitride layer, are more down-shifted by the GW corrections than the double-degenerate valence bands (HOMO-1), located on $WS_2$, which then become the new VBM. In other words, the valence band which is the highest in DFT (HOMO) becomes the HOMO-1 in GW, and the band which in DFT is  HOMO-1, becomes the new HOMO in GW. This band exchange is due to two reasons: first  the different nature of the top valence bands belonging to the AlN and WS$_2$, the former are mainly formed from $p$ orbitals of the N atom while the latter originate from the $d$ orbitals of the W atom, and so they acquire a different GW correction as it happens, often, in molecular systems\cite{faber2012electron}. Second it is the different screening felt by the electrons in AlN or in WS2, in fact, the screening is less effective in AlN (it has a larger gap), and this causes a larger quasi-particle correction on the AlN states with respect to WS2 states.
Therefore, while at the DFT level a type II heterostructure is obtained, 
the GW corrections modify this picture and give a type I system, where both top valence and bottom conduction bands belong to the WS$_2$ subsystem. \\
In this context it is worth to point out that the type I band offset can be obtained only considering both SOC and many-body effects\cite{marsili2021spinorial}. 
Indeed in a recent work by Yeganeh et al. \cite{Yeganeh2022} a type II band alignment at G${_0}$W${_0}$ level but without inclusion of  SOC was found. 

   
  
    \subsection{Optical Properties of AlN/WS$_2$}
    
\begin{figure}[ht]
        \centering
        \includegraphics[width=0.5\textwidth]{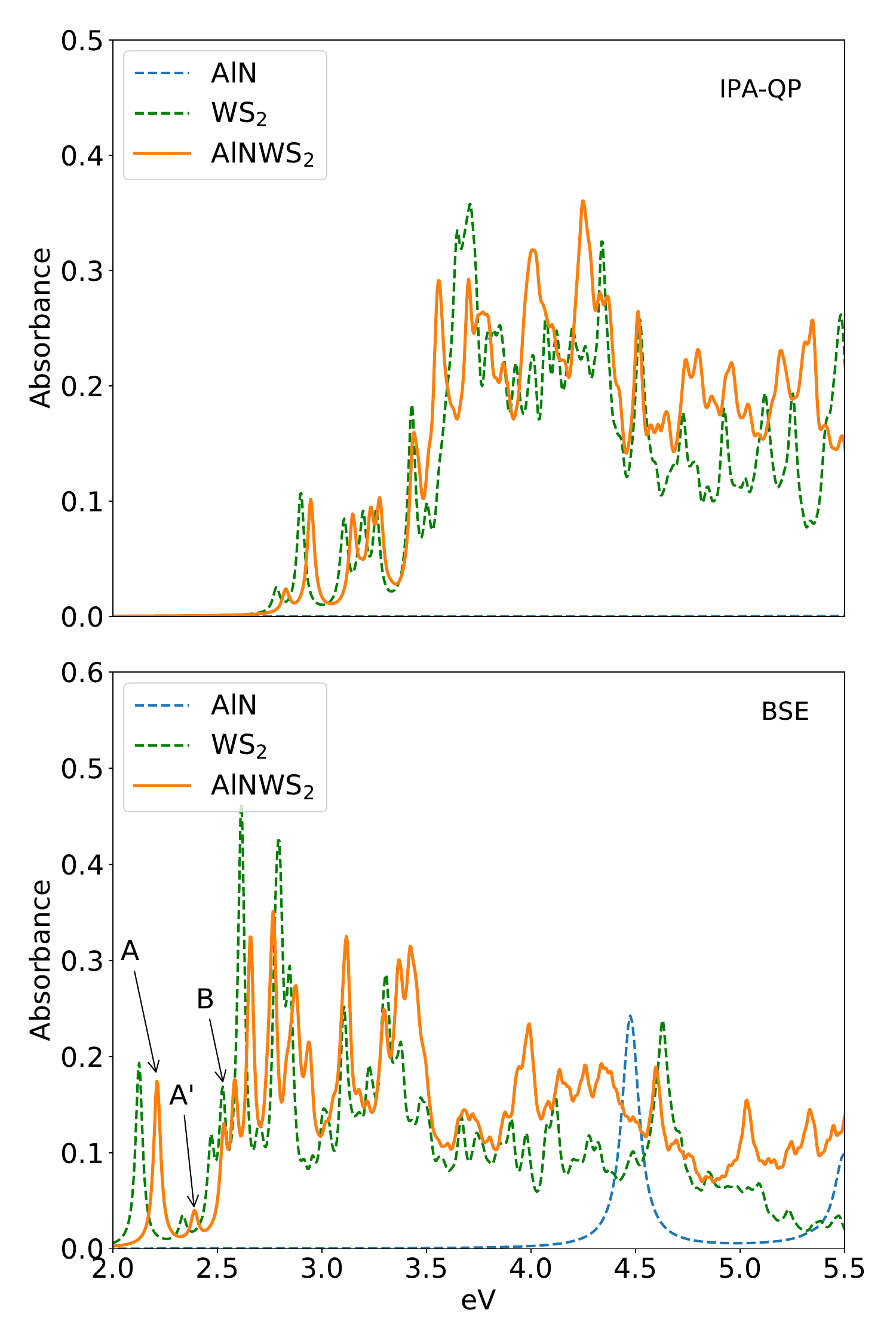}
	      \caption{Absorbance of  AlN/WS$_2$ heterostructure, and of AlN and WS$_2$ isolated, calculated within the independent-quasi-particle approach G$_0$W$_0$ level  (top panel) and Bethe-Salpeter Equation (bottom panel) 
	      \label{fig:AlN/WS2_RPA_opticalcond}}
      \end{figure}

    The optical response of the HT is investigated at two different levels of theory: within the independent-QP approach at the   G$_0$W$_0$ level, and  with the inclusion of local-fields and excitonic effects by solving the BSE. In Fig. \ref{fig:AlN/WS2_RPA_opticalcond} we present the in-plane optical  absorbance for the two cases.
    
In order to understand the optical properties of the heterostrucure, we start with the study of the two separated layers.

  The optical properties of free standing AlN, in the energy range we are interested in, are dictated by a single excitonic peak at 4.75~eV two times degenerate, see the bottom panel of Fig.~\ref{fig:AlN/WS2_RPA_opticalcond}, dashed blue line. 
This direct exciton is strongly bound, has large oscillator strength and a very small radius (see Tab.~\ref{tab:2Dexc}). 
Because of the large GW direct  gap (6.27 eV), no optical response is visible at the independent-QP level in the range here considered (0-5.5 eV, see Fig.~\ref{fig:AlN/WS2_RPA_opticalcond} upper panel).

\begin{figure}[!ht]
        \centering
        \includegraphics[width=0.475\textwidth]{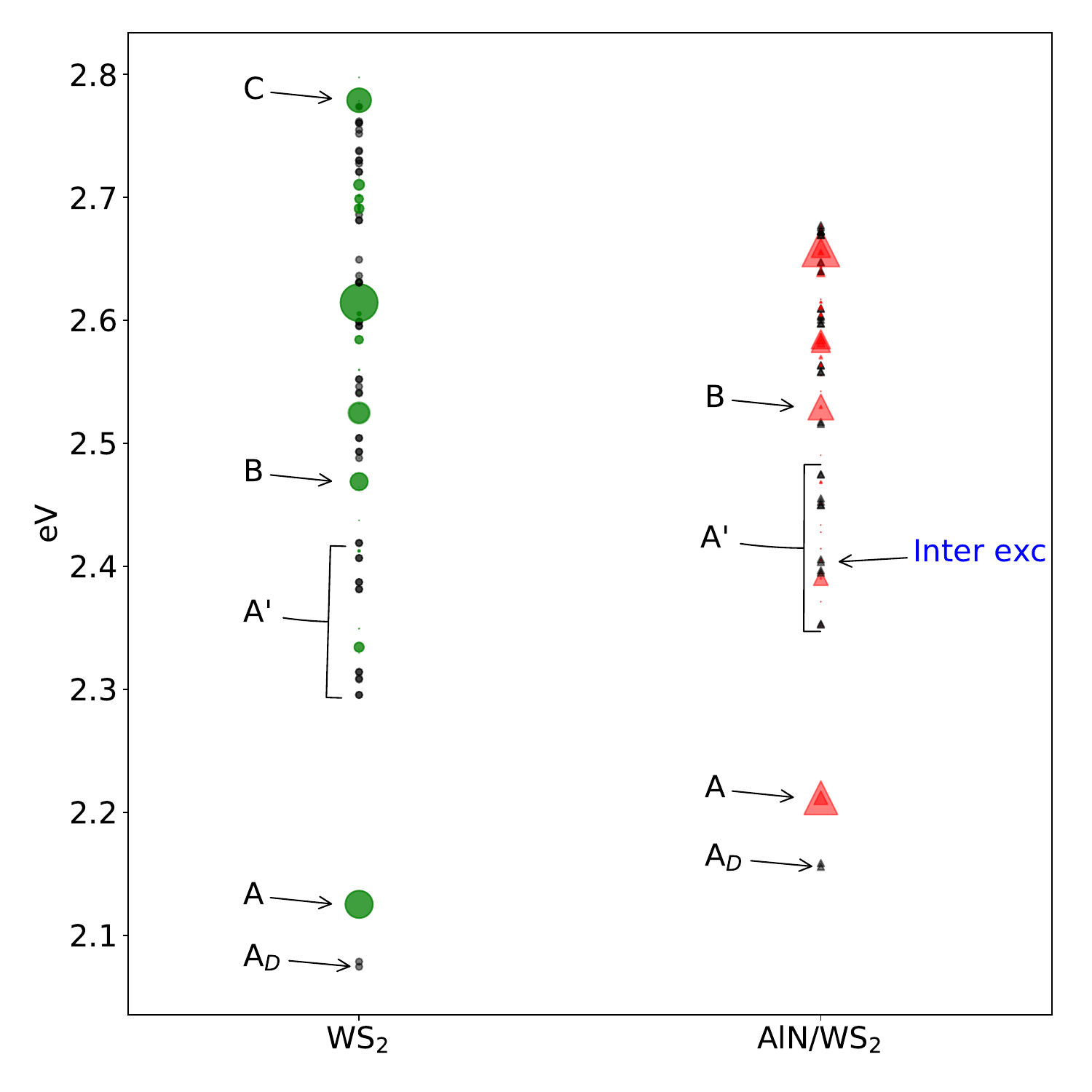}%
        \caption{Schematic representation of excitons level in WS$_2$ layer and AlN/WS$_2$ bilayer. Here we report exciton at zero momentum, the ones responsible for the optical absorption. The size of the dots is proportional to the dipole of the different excitons. Dark states are represented by black dots. In the AlN/WS$_2$ case we also report the position of the first inter-layer exciton. 
		 \label{fig:AlN/WS2_excitons}}
      \end{figure}

The optical response of WS$_2$ (green dashed line in Fig.\ref{fig:AlN/WS2_RPA_opticalcond}) is 
more complex.  The schematic representation of the WS$_2$ excitonic levels is shown in Fig.~\ref{fig:AlN/WS2_excitons}. The lowest exciton at 2.075~eV, 
is the spin-forbidden dark and double degenerate, 
generally called A$_D$.
The first bright, also double degenerate, exciton $A$ is at 2.125~eV.
Then, the B exciton, due to spin-orbit split bands,  is at 2.46~eV.
Between the A and B, a series of small peaks, called A', coming from transitions near $K,K'$ and with a very small dipole matrix elements, are present.
These results are in agreement with theoretical predictions and experimental data\cite{mak2010atomically},\cite{Ugeda2014},\cite{Zhu2015}, \cite{palummo2015exciton}. 
At higher energy other excitons (normally called C and D)  due to transitions in the band-nesting $\Gamma-K$ region, are visible. 
According to the literature\cite{Molinas2016}, they are more difficult to converge and are also more affected by the electron-phonon interaction which is not included in the present work. 
 The WS$_2$ exciton levels (see Fig.~\ref{fig:AlN/WS2_excitons}) are in good agreement with previous calculation of Marsili et al.\cite{marsili2021spinorial} and the small differences are due to the different pseudo-potentials used in the calculations. \\
We now move to the analysis of optical properties of the AlN/WS$_2$ bilayer. 
     As for WS$_2$, the inclusion of the attractive e-h interaction  on top of the GW calculations moves back the absorption spectrum toward lower energy and produce new peaks due to strongly bound excitons (orange solid lines in Fig. \ref{fig:AlN/WS2_RPA_opticalcond}).
     Hence, the AlN/WS$_2$ HT shows substantial adsorption from visible to deep UV light. Its spectrum is very similar to the WS$_2$ free standing one, but slightly blue shifted.

      
      In order to understand the origin of the peaks and of the blue shift, we analyze the bands involved in the transitions.
      
      Since, as mentioned,  there is an interchange of the ordering of  the GW bands, in the following, when we talk about the electronic bands involved in the excitonic transitions, we  refer to the original DFT order. This is to avoid confusion in the analysis of the excitons.
      The first direct exciton at 2.15 eV, is dark and double degenerate, and it is mainly  due  to the transition between the HOMO-1  and the LUMO at the {\bf K}  point, with a binding energy of 0.64 eV. The first visible exciton, also twice degenerate, appears at 2.21 eV and  is characterized by a binding energy E$_b$ = 0.59 eV.  It involves the same bands of the two lowest dark excitons. The exciton binding energy is smaller than that of the isolated WS$_2$ (0.64 eV) due to the presence of the AlN layer that increases the screening of the electron-hole interaction. This reduction of the binding energy contributes to the blue shift of the excitons.
      The peak around 2.65 eV
       belongs to transition at K between the third band below the highest valence band (HOMO-2) and the second conduction band (LUMO+1), and it also has double degeneration. An analysis of the first ten excitons (most of them degenerate) of the AlN/WS$_2$ in terms of valence-conduction transitions shows that all correspond to excitation involving just  WS$_2$ states. 
       In fact the lowest inter-layer 
       states  between AlN and WS$_2$ appear at an energy corresponding to the  $A'$ states around  2.40~eV. These states are not visible in the spectra due to their very small dipole moment. 
      By comparing the AlN/WS$_2$ spectra with those   of the isolated AlN and WS$_2$ (see Fig.\ref{fig:AlN/WS2_RPA_opticalcond}) we can conclude that the dominant contribution comes from WS$_2$.  AlN contribution is limited to the high energy region due to the large optical gap of about 4.6 eV\cite{prete2020giant} but, indirectly, affects the WS$_2$ through the increase of  the screening and, as we will discuss in the next section, through the lattice mismatch.      
      \subsubsection{Effect of the substrate \label{subsec:strain}}
      
      
      As shown in Fig. \ref{fig:AlN/WS2_RPA_opticalcond}, lower panel, the first exciton in AlN/WS$_2$ is blue-shifted by about 0.1 eV with respect to the first exciton of isolated  WS$_2$. This is a puzzling result since several experiments and theoretical calculations show an opposite trend,\cite{latini2015excitons,Ugeda2014}: an increasing of screening tends to red-shift the optical spectra. But care must be taken when directly comparing two spectra. Indeed, several (interconnected) effects are responsible of the final change in the spectra (Fig.~\ref{fig:AlN/WS2_RPA_opticalcond}): the variation of the screening, the variation of the  direct gap and the change in the lattice constant. In order to test and disentangle these two effects, we also simulate an AlN/WS$_2$ bilayer forcing the lattice constant to be the same of the isolated WS$_2$ monolayer (a=$3.16$ \r{A}). In this case we find that the direct gap reduces of $-0.05$~eV with respect to the isolated WS$_2$,  due to the additional screening generated by the presence of the AlN layer. On the other hand, as we have discussed before, see Tab.~\ref{fig:configurations}, the full relaxation of the bilayer induces a small compressive strain on WS$_2$ (a=$3.142$ \r{A}) in the HT. This increases the direct gap of $0.09~eV$ with respect to the unrelaxed bilayer (constrained to have a=3.16 \AA) and leaves the binding energy unchanged. This means that the variation of the lattice constant induces a change in the gap larger than the one due to 
 AlN layer screening.
     The small strain applied has then essentially no effect on the excitonic binding energy, consistently with other works\cite{lechifflart2022excitons}. 
     To summarize,
    this explains why the fully relaxed heterostructure has a larger GW gap than the isolated WS$_2$, even in the presence of a larger screening. 
    This finding, combined with the reduction of the excitonic binding energy due to the substrate (AlN), explains the blue-shift of the first exciton in the Fig.~\ref{fig:AlN/WS2_RPA_opticalcond}.\\ 

      
    \subsubsection{Exciton model}  
 Finally, it is worth to notice that a simple 2D excitonic model based on the Rytova-Keldysh approach \cite{rytova, Keldysh:1979:JETPL,pulci2015excitons,cudazzo2011dielectric} would give reasonable results, predicting  the trend of the excitonic binding energies close to the ab-initio values.

Within this model, the Hamiltonian which describe the interaction of an electron-hole pair in a homogeneous 2D sheet  with parabolic bands is given by
\begin{equation}\label{RK}
\left\{E_g   -\frac{\hbar^2}{2\mu}\bm{\nabla}^2_{\rho}    +\hat{W}({\rho})\right\}\phi_0({\rho})=E_0\phi_0({\rho})
\end{equation}
with $\mu$ exciton reduced mass and $\hat{W}({\rho})$ the statically screened electron-hole attraction potential:
\begin{equation}
\label{RK_pot}
W(\rho)=-\frac{\pi e^2}{2\rho_0}\left[H_0\left(\frac{\rho}{\rho_0}\right)-N_0\left(\frac{\rho}{\rho_0}\right)\right].
\end{equation}
Here, $H_0$ is the Struve function, $N_0$ the Neumann Bessel function of the second kind, and $\rho_0$ is the screening radius: $\rho_0=2\pi\alpha_{2D}$ with $\alpha_{2D}$ the static electronic sheet polarizability.

 Although   the error is quite large (see Tab.~\ref{tab:2Dexc}) this method confirms to be a cheap, fast way to obtain qualitative results. The simplicity of this method relies on the fact that the effective masses and polarizabilities are the only needed ingredients, and both can be easily obtained within DFT.

          \begin{table}[!ht]
      \centering
      \begin{tabular}{|c|c|c|c|c|}
      \hline \hline
            &  m$^*_h$/m$^*_e$ (m$_e$)& Re $\alpha_{2D}$ (a.u.) & E$_b$ (eV) & r$_{exc}$ (\AA) \\ \hline  
AlN         &  1.5/0.66   & 2.2 & 1.79(2.3)& 3.4 \\  \hline
WS$_2$      &  0.42/0.44   & 13.0 & 0.64(0.5)& 11.4 \\  \hline             
AlNWS$_2$   &  0.41/0.38   & 15.6 & 0.59(0.4) & 12.8 \\  \hline \hline
      \end{tabular}
		  \caption{The excitonic binding energies of the first bright excitons calculated within the BSE. In parenthesis, the values obtained within the 2D excitonic model are reported. In the last column we report also the excitonic radius $r_{exc}$ obtained using the same model.
      }\label{tab:2Dexc}
      \end{table}

\section{Conclusions}

     In summary, we have presented a study on a novel AlN/WS$_2$ vdW heterostructure. Its negative formation energy as well as its small lattice mismatch between the constituent monolayers suggests that it  should  be possible to synthesize it.  The analysis of the band alignment and decomposed partial charge density of the heterostructure shows that a calculation at the DFT level is not sufficient to identify the character of the heterostructure. The GW inverts the band order transforming the heterostructure from type II to type I. The subsequent Bethe-Salpeter calculation of optical excitations confirms this result. All the lowest optical excition, dark or bright, belong to the WS$_2$ monolayer part.
      These results suggest that a simple identification of heterostructures by means of local or semi-local functional in DFT may fail when compared with more advanced methods as GW plus Bethe-Salpeter or hybrid functionals.

Finally, the optical spectrum of the heterostructure is very similar to the one of the isolated WS$_2$, with small differences arising from the presence of the substrate, which acts through the strain and through the screening. We demonstrate that the AlN substrate modifies only slightly the electronic and optical properties of WS$_2$ and therefore could be employed as an excellent alternative insulating substrate, acting as gate dielectric material, to the most common used h-BN.\cite{prete2020giant}

\section{acknowledgments}
The research leading to these results has  received funding from the European Union Seventh Framework Program under grant agreement no. 7
85219 Graphene Core2. C.A. acknowledges A. Saul and K. Boukari for the management of the computer cluster \emph{Rosa}. CPU time was also granted by CINECA (ISCRA-B and ISCRA-C) and CRESCO ENEA HPC centers.  This publication is based upon work from COST Action TUMIEE CA17126, supported by COST (European Cooperation in Science and Technology). O.P. and M.P. acknowledge financial funding from the EU MSCA-RISE project  DiSeTCom  (HORIZON2020, GA 823728).
M.P. acknowledges  Tor Vergata University Project 2021 TESLA. O.P. acknowledges fundings from PRIN 2020 "PHOTO".

\section{Abbreviations}
The following abbreviations are used in this manuscript:\\

\noindent 
\begin{tabular}{@{}ll}
ALD & Atomic layer deposition \\
BSE & Bethe-Salpeter Equation\\
CBM & Conduction band minimum\\
DFT & Density Functional Theory\\
GGA & Generalized gradient approximation\\
HOMO & highest occupied molecular orbital \\
HT & Heterostructure\\
LED & Light emitting device \\ 
ML & Monolayer \\ 
MBPT & Many-body perturbation theory \\ 
LUMO & Lowest unoccupied molecular orbital\\
PDOS & Projected density of states \\
PV & Photo-voltaic \\ 
SOC & Spin orbit coupling \\
TMD & transition metal dichalcogenides \\ 
VBM & Valence band maximum 
\end{tabular}

\bibliography{biblio_stella}

\end{document}